\documentclass[a4paper,english,runningheads]{llncs}
\usepackage[T1]{fontenc}
\usepackage[latin9]{inputenc}
\usepackage{mathrsfs}
\usepackage{amsmath}
\usepackage{amssymb}

\makeatletter

\usepackage[all]{xy}

\usepackage{cite}

\usepackage{babel}

\usepackage{babel}

\makeatother

\usepackage{babel}
\begin{document}

\title{Contextuality-by-Default: A Brief Overview of Ideas, Concepts, and
Terminology }

\titlerunning{Contextuality-by-Default}

\author{Ehtibar N. Dzhafarov\textsuperscript{1}, Janne V. Kujala\textsuperscript{2},
and Victor H. Cervantes\textsuperscript{1}}

\authorrunning{E. N. Dzhafarov, J. V. Kujala, V. H. Cervantes }

\institute{\textsuperscript{1}Purdue University\\
 ehtibar@purdue.edu\\
 $\,$\\
 \textsuperscript{2}University of Jyväskylä\\
 jvk@iki.fi}

\toctitle{Contextuality-by-Default: Logic, Concepts, Terminology}

\tocauthor{E. N. Dzhafarov, J. V. Kujala, V. H. Cervantes}
\maketitle
\begin{abstract}
This paper is a brief overview of the concepts involved in measuring
the degree of contextuality and detecting contextuality in systems
of binary measurements of a finite number of objects. We discuss and
clarify the main concepts and terminology of the theory called ``contextuality-by-default,''
and then discuss generalizations of the theory to arbitrary systems
of arbitrary random variables.

\end{abstract}

\section{Introduction}

\subsection{On the name of the theory}

The name ``contextuality-by-default'' should not be understood as
suggesting that any system of measurements is contextual, or contextual
unless proven otherwise. The systems are contextual or noncontextual
depending on certain criteria, to be described. The name of the theory
reflects a philosophical position according to which every random
variable's identity is inherently contextual, i.e., it depends on
all conditions under which it is recorded, whether or not there is
a way in which these conditions could affect the random variable physically.
Thus, in the well-known EPR-Bell paradigm, Alice and Bob are separated
by a space-like interval that prevents Bob's measurements from being
affected by Alice's settings; nevertheless, Bob's measurements should
be labeled by both his own setting and by Alice's setting; and as
the latter changes with the former fixed, the identity of the random
variable representing Bob's measurement changes ``by default.''
One does not have to ask ``why.'' Bob's measurements under two different
settings by Alice have no probabilisitic relation to each other; they
possess no joint distribution. Therefore one cannot even meaningfully
ask the question of whether the two may be ``always equal to each
other.'' The questions one can ask meaningfully are all about what
joint distributions can be \emph{imposed} (in a well-defined sense)
on the system in a way consistent with observations. A system is contextual
or noncontextual (or, as we say, has a noncontextual description)
depending on the answers to these questions. Thus, the famous Kochen-Specker
demonstration of contextuality is, from this point of view, a reductio
ad absurdum proof that measurements of a spin in a fixed direction
made under different conditions cannot be imposed a joint distribution
upon, in which these measurements would be equal to each other with
probability 1.

\subsection{Notation}

We use capital letters $A,B,\ldots,Q$ to denote sets of ``objects''
(properties, quantities) being measured, and the script letter $\mathscr{C}$
to denote a collection of such sets. We use capital letters $R,S$,
and $T$ to denote the measurements (random variables), and the Gothic
letter $\mathfrak{R}$ to denote sets of random variables that do
not possess a joint distribution.

\section{Contextuality-by-Default at a Glance}

The following is an overview of the main concepts and definitions
of the contextuality-by-default theory. This is not done at a very
high level of generality, in part in order not to be too abstract,
and in part because the criterion and measure of contextuality have
been developed in detail only for a broad subclass of finite sets
of binary measurements. Thus, the notion of a context given below
in terms of subsets of measured objects is limited, but it is intuitive,
and a way to generalize it is clear (Section \ref{sec:Generalizations-Beyond-Binary}).
The definition of maximally connected couplings is given for binary
($\pm1$) measurements only, and reasonable generalizations here may
not be unique. In Section \ref{sec:Generalizations-Beyond-Binary}
we discuss one, arguably most straightforward way of doing this.

\subsection{Measurements are labeled contextually}

There is a set $Q$ of ``\emph{objects}'' we want to measure. For
whatever reason, we cannot measure them ``\emph{all at once}'' (the
meaning of this is not necessarily chronological, as explained in
Section \ref{sec:Conceptual-and-Terminological}). Instead we define
a collection of subsets of $Q$, 
\begin{equation}
\mathscr{C}=\left\{ A\subset Q,B\subset Q,\ldots\right\} ,
\end{equation}
and measure the objects ``\emph{one subset at a time}.'' We call
these subsets of objects \emph{contexts}. Different contexts may overlap.
This definition has limited applicability, and we discuss a general
definition in Section \ref{sec:Generalizations-Beyond-Binary}. For
now we consider only finite sets $Q$ (hence finite collections of
finite contexts).

The \emph{measurement outcome} of each object $q$ (from the set $Q$)
in each context $C$ (from the collection $\mathscr{C}$) is a random
variable, and we denote it $R_{q}^{C}$ (with $q\in C\in\mathscr{C}$).
This is called \emph{contextual labeling} of the measurement outcomes.
It ensures that the collection 
\begin{equation}
\left\{ R_{q}^{A}\right\} _{q\in A},\left\{ R_{q}^{B}\right\} _{q\in B},\ldots
\end{equation}
for all $A,B,\ldots$ comprising $\mathscr{C}$, are pairwise disjoint,
as no two random variables taken from two different members of the
collection have the same superscript (whether or not they have the
same subscript).

\subsection{Measurements in different contexts are stochastically unrelated}

We call $R^{C}=\left\{ R_{q}^{C}\right\} _{q\in C}$ for every $C\in\mathscr{C}$
a \emph{bunch} (of random variables). The random variables within
a bunch are \emph{jointly distributed}, because of which we can consider
each bunch as a single (multicomponent) random variable. If $q,q'\in C\in\mathscr{C}$,
we can answer questions like ``what is the correlation between $R_{q}^{C}$
and $R_{q'}^{C}$?''. However, if $q\in C$, $q'\in C'$, and $C\not=C'$,
then we cannot answer such questions: $R_{q}^{C}$ and $R_{q'}^{C'}$
belong to different bunches and do not have a joint distribution.
We say that they are \emph{stochastically unrelated}.

\subsection{\label{sub:All-possible-couplings}All possible couplings for all
measurements }

Consider now the (necessarily disjoint) union of all bunches 
\begin{equation}
\mathfrak{R}=\bigcup_{C\in\mathscr{C}}R^{C}=\bigcup_{C\in\mathscr{C}}\left\{ R_{q}^{C}\right\} _{q\in C},
\end{equation}
i.e., the set of all measurements contextually labeled. The use of
the Gothic font is to emphasize that this set is \emph{not} a multicomponent
random variable: except within bunches, its components are not jointly
distributed. We call $\mathfrak{R}$ a \emph{system} (of measurements).

Now, we can be interested in whether and how one could \emph{impose
a joint distribution} on $\mathfrak{R}$. To impose a joint distribution
on $\mathfrak{R}$ means to find a set of jointly distributed random
variables $S=\left\{ S_{q}^{C}\right\} _{q\in C\in\mathscr{C}}$ such
that, for every $C\in\mathscr{C}$, 
\begin{equation}
S^{C}=\left\{ S_{q}^{C}\right\} _{q\in C}\sim\left\{ R_{q}^{C}\right\} _{q\in C}=R^{C}.
\end{equation}
The symbol $\sim$ means ``has the same distribution as.'' Note
that 
\begin{equation}
S=\left\{ S^{C}\right\} _{C\in\mathscr{C}}=\left\{ S_{q}^{C}\right\} _{q\in C\in\mathscr{C}}
\end{equation}
is a single (multicomponent) random variable, and in probability theory
$S$ is called a \emph{coupling} for (or of) $\mathfrak{R}$. Any
subset of the components of $S$ is its \emph{marginal}, and $S^{C}$
is the marginal of $S$ whose components are labeled in the same way
as are the components of the bunch $R^{C}$.

If no additional constraints are imposed, one can always find a coupling
$S$ for any union of bunches. For instance, one can always use an
\emph{independent coupling}: create a \emph{copy} $S^{C}$ of each
bunch $R^{C}$ (i.e., an identically labelled and identically distributed
set of random variables), and join them so that they are stochastically
independent. The set $S=\left\{ S^{C}\right\} _{C\in\mathscr{C}}$
is then jointly distributed. The existence of a coupling \emph{per
se} therefore is not informative.

\subsection{\label{sub:Connections-and-their}Connections and their couplings}

Let us form, for every object $q$, a set of random variables 
\begin{equation}
\mathfrak{R}_{q}=\left\{ R_{q}^{C}\right\} _{C\in\mathscr{C}},
\end{equation}
i.e., all random variables measuring the object $q$, across all contexts,
We call this set, which is not a random variable, a \emph{connection}
(for $q$). Let us adopt the convention that if a context $C$ does
not contain $q$, then $R_{q}^{C}$ is not defined and does not enter
in $\mathfrak{R}_{q}$. 

A system is called \emph{consistently connected} if, for every $q\in Q$
and any two contexts $C,C'$ containing $q$, 
\begin{equation}
R_{q}^{C}\sim R_{q}^{C'}.
\end{equation}
Otherwise a system is called (strictly) \emph{inconsistently connected}.
Without the adjective ``strictly,'' inconsistent connectedness means
that the equality above is not assumed, but it is then not excluded
either: consistent connectedness is a special case of inconsistent
connectedness.

One possible interpretation of strictly inconsistent connectedness
is that the conditions under which a context is recorded may physically
influence (in some cases one could say, ``signal to'') the measurements
of the context members. Another possibility is that a choice of context
may introduce biases in how the objects are measured and recorded.

\subsection{Maximally connected couplings for binary measurements}

Every coupling $S$ for $\mathfrak{R}$ has a marginal $S_{q}=\left\{ S_{q}^{C}\right\} _{C\in\mathscr{C}}$
that forms a coupling for the connection $\mathfrak{R}_{q}$. We can
also take $\mathfrak{R}_{q}$ for a given $q$ in isolation, and consider
all its couplings $T_{q}=\left\{ T_{q}^{C}\right\} _{C\in\mathscr{C}}$.
Clearly, the set of all $S_{q}$ extracted from all possible couplings
$S$ for $\mathfrak{R}$ is a subset of all possible couplings $T_{q}$
for $\mathfrak{R}_{q}$.

Let us now confine the consideration to \emph{binary measurements}:
each random variable in the system has value $+1$ or $-1$. In Section
\ref{sec:Generalizations-Beyond-Binary} we will discuss possible
generalizations.

A coupling $T_{q}$ for a connection $\mathfrak{R}_{q}$ is called
\emph{maximal} if, given the expected values $\left\langle R_{q}^{C}\right\rangle $
for all $C$, the value of 
\begin{equation}
\mathsf{eq}\left(T_{q}\right)=\Pr\left[T_{q}^{C}=1:C\in\mathscr{C}\right]+\Pr\left[T_{q}^{C}=-1:C\in\mathscr{C}\right]\label{eq:EQ-type probs}
\end{equation}
is the largest possible among all couplings for $\mathfrak{R}_{q}$
(again, $R_{q}^{C}$ and $T_{q}^{C}$ are not defined and are not
considered if $C$ does not contain $q$).

Let us denote 
\begin{equation}
\max\mathsf{eq}\left(\mathfrak{R}_{q}\right)=\max_{\substack{\textnormal{all possible}\\
\textnormal{couplings \ensuremath{T_{q}}for }\mathfrak{R}_{q}
}
}\mathsf{eq}\left(T_{q}\right).\label{eq:meqR_q}
\end{equation}
It follows from a general theorem mentioned in Section \ref{sec:Generalizations-Beyond-Binary}
that this quantity is well-defined for all systems, i.e., that the
supremum of $\mathsf{eq}\left(T_{q}\right)$ is attained in some coupling
$T_{q}$. Clearly, for consistently connected systems $\max\mathsf{eq}\left(\mathfrak{R}_{q}\right)=1$
(the measurements can be made ``perfectly correlated''). For (strictly)
inconsistently connected systems, $\max\mathsf{eq}\left(\mathfrak{R}_{q}\right)$
is always well-defined, and it is less than $1$ for some $q$. It
may even be zero: for $\pm1$ variables this happens when the $\mathfrak{R}_{q}$
contains two measurements $R_{q}^{A}$ and $R_{q}^{B}$ such that
$\Pr\left[R_{q}^{A}=1\right]=1$ and $\Pr\left[R_{q}^{B}=1\right]=0$.

\subsection{Definition of contextuality}

Consider again a coupling $S$ for the entire system $\mathfrak{R}$,
and for every $q\in Q$, extract from $S$ the marginal $S_{q}$ that
forms a coupling for the connection $\mathfrak{R}_{q}$. \medskip{}

\textbf{Central Concept.} If, for every $q\in Q$, 
\begin{equation}
\mathsf{eq}\left(S_{q}\right)=\max\mathsf{eq}\left(\mathfrak{R}_{q}\right),
\end{equation}
(i.e., if every marginal $S_{q}$ in $S$ is a maximal coupling for
$\mathfrak{R}_{q}$) then the coupling $S$ for $\mathfrak{R}$ is
said to be \emph{maximally connected}. \medskip{}

\noindent Intuitively, in this case the measurements can be imposed
a joint distribution upon in which the measurements $R_{q}^{C}$ of
every object $q$ in different contexts $C$ are maximally ``correlated,''
i.e., attain one and the same value with the maximal probability allowed
by their observed individual distributions (expectations). \medskip{}

\textbf{Main Definition.} A system $\mathfrak{R}$ is said to be \emph{contextual}
if no coupling $S$ of this system is maximally connected. Otherwise,
a maximally connected coupling of the system (it need not be unique
if it exists) is said to be this system's \emph{noncontextual description}
(or, as a terminological variant, \emph{maximally noncontextual description}).

\medskip{}

For consistently connected systems this definition is equivalent to
the traditional understanding of (non)contextuality. According to
the latter, a system has a noncontextual description if and only if
there is a coupling for the measurements \emph{labeled} \emph{noncontextually}.
The latter means that all random variables $R_{q}^{C}$ within a connection
are treated as being equal to each other with probability 1.

\subsection{Measure and criterion of contextuality}

If (and only if) a system $\mathfrak{R}$ is contextual, then for
every coupling $S$ there is at least one $q\in Q$ such that $\mathsf{eq}\left(S_{q}\right)<\mathsf{\max\mathsf{eq}}\left(\mathfrak{R}_{q}\right)$.
This is equivalent to saying that a system is contextual if and only
if for every coupling $S$ of it, 
\begin{equation}
\sum_{q\in Q}\mathsf{eq}\left(S_{q}\right)<\sum_{q\in Q}\max\mathsf{eq}\left(\mathfrak{R}_{q}\right).\label{eq:ineq. for contextual}
\end{equation}
Define 
\begin{equation}
\max\mathsf{eq}\left(\mathfrak{R}\right)=\max_{\substack{\textnormal{all couplings}\\
S\textnormal{ for }\mathfrak{R}
}
}\left(\sum_{q\in Q}\mathsf{eq}\left(S_{q}\right)\right).\label{eq:meqR}
\end{equation}
In this definition we assume that this maximum exists, i.e., the supremum
of the sum on the right is attained in some coupling $S$. (This is
likely to be true for all systems with finite $Q$ and binary measurements,
but we only have a formal proof of this for the cyclic systems considered
below.) Then (\ref{eq:ineq. for contextual}) is equivalent to 
\begin{equation}
\max\mathsf{eq}\left(\mathfrak{R}\right)<\sum_{q\in Q}\max\mathsf{eq}\left(\mathfrak{R}_{q}\right),\label{eq:criterion}
\end{equation}
which is a \emph{criterion of contextuality} (necessary and sufficient
condition for it). Moreover, it immediately leads to a natural \emph{measure
of contextuality}: 
\begin{equation}
\mathsf{cntx}\left(\mathfrak{R}\right)=\sum_{q\in Q}\max\mathsf{eq}\left(\mathfrak{R}_{q}\right)-\max\mathsf{eq}\left(\mathfrak{R}\right).\label{eq:measure}
\end{equation}
Written \emph{in extenso} using (\ref{eq:meqR_q}) and (\ref{eq:meqR}),
\begin{equation}
\mathsf{cntx}\left(\mathfrak{R}\right)=\sum_{q\in Q}\max_{\substack{\textnormal{all possible}\\
\textnormal{couplings \ensuremath{T_{q}}for }\mathfrak{R}_{q}
}
}\mathsf{eq}\left(T_{q}\right)-\max_{\substack{\textnormal{all couplings}\\
S\textnormal{ for }\mathfrak{R}
}
}\sum_{q\in Q}\mathsf{eq}\left(S_{q}\right),
\end{equation}
where, one should recall, $S_{q}$ is the marginal of $S$ that forms
a coupling for $\mathfrak{R}_{q}$. We can see that the minuend and
subtrahend in the definition of $\mathsf{cntx}\left(\mathfrak{R}\right)$
differ in order of the operations $\max$ and $\sum_{q\in Q}$; and
while in the minuend the choice of couplings $T_{q}$ for $\mathfrak{R}_{q}$
is unconstrained, in the subtrahend the choice of couplings $S_{q}$
for $\mathfrak{R}_{q}$ is constrained by the requirement that it
is a marginal of the coupling for the entire system $\mathfrak{R}$.

\section{The history of the contextuality-by-default approach}

A systematic realization of the idea of contextually labeling a system
of measurements $\mathfrak{R}$, considering all possible couplings
$S$ for it, and characterizing it by the marginals $S_{q}$ that
form couplings for the connections $\mathfrak{R}_{q}$ of the system
was developing through a series of publications \cite{DK2013PLOS,DK2014LNCSQualified,DK2014NoForcing,DK2014PAMS,DK2014PLOSconditionalization}.
The idea of maximally connected couplings as the central concept for
contextuality in consistently connected systems was proposed in Refs.
\cite{DK2014Advances,DK2014Scripta,deBarros} and then generalized
to inconsistently connected systems \cite{DK2014Arxiv1,DK2014CbDArxiv2}.

In the latter two references the measure of contextuality (\ref{eq:measure})
and the criterion of contextuality (\ref{eq:criterion}) were defined
and computed for simple QM systems (cyclic systems of rank 3 and 4,
as defined below). Later we added to this list cyclic systems of rank
5, and formulated a conjecture for the measure and criterion formulas
for cyclic systems of arbitrary rank \cite{DKL2015FooP}.

In Refs. \cite{DKL2015FooP,KDL2015PRL} the contextuality-by-default
theory is presented in its current form. The conjecture formulated
in Ref. \cite{DKL2015FooP} was proved in Ref. \cite{KD2015}.

A cyclic system (with binary measurements) is defined as one involving
$n$ ``objects'' ($n$ being called the\emph{ rank} of the system)
measured two at a time, 
\begin{equation}
\left(q_{1},q_{2}\right),\left(q_{2},q_{3}\right),\ldots,\left(q_{n-1},q_{n}\right),\left(q_{n},q_{1}\right).
\end{equation}
For $i=1,\ldots,n$, the pair $\left(q_{i},q_{i\oplus1}\right)$ forms
the context $C_{i}$ ($\oplus$ standing for circular shift by 1).
Each object $q_{i}$ enters in precisely two consecutive contexts,
$C_{i\ominus1}$ and $C_{i}$. Denoting the measurement of $q_{i}$
in context $C_{j}$ by $R_{i}^{j}$, we have the system represented
by bunches $R^{i}=\left(R_{i}^{i},R_{i\oplus1}^{i}\right)$ and connections
$\mathfrak{R}_{i}=\left(R_{i}^{i\ominus1},R_{i}^{i}\right)$.

The formula for the measure of contextuality conjectured in Ref. \cite{DKL2015FooP}
and proved in Ref. \cite{KD2015} is 
\begin{equation}
\mathsf{cntx}\left(\mathfrak{R}\right)=\frac{1}{2}\max\left\{ \begin{array}{l}
\textnormal{s}_{\textnormal{odd}}\left(\left\langle R_{i}^{i}R_{i\oplus1}^{i}\right\rangle :i=1,\dots,n\right)-\sum_{i=1}^{n}\left|\left\langle R_{i}^{i}\right\rangle -\left\langle R_{i}^{i\ominus1}\right\rangle \right|-\left(n-2\right)\\
0
\end{array}\right..\label{eq:measure cyclic}
\end{equation}
The function $\textnormal{s}_{\textnormal{odd}}$ is defined for an
arbitrary set of argument $x_{1},\ldots,x_{k}$ as 
\begin{equation}
\textnormal{s}_{\textnormal{odd}}\left(x_{1},\ldots,x_{k}\right)=\max\left(\pm x_{1}\pm\ldots\pm x_{k}\right),
\end{equation}
where the maximum is taken over all assignments of $+$ and $-$ signs
with an odd number of $-$'s. The criterion of contextuality readily
derived from (\ref{eq:measure cyclic}) is: the system is contextual
if ands only if 
\begin{equation}
\textnormal{s}_{\textnormal{odd}}\left(\left\langle R_{i}^{i}R_{i\oplus1}^{i}\right\rangle :i=1,\dots,n\right)>\left(n-2\right)+\sum_{i=1}^{n}\left|\left\langle R_{i}^{i}\right\rangle -\left\langle R_{i}^{i\ominus1}\right\rangle \right|.\label{eq:criterion cyclic}
\end{equation}
For consistently connected systems, the sum on the right vanishes,
and we can derive the traditional formulas for Legget-Garg ($n=3$),
EPR/Bell ($n=4$), and Klyachko-Can-Binicioglu-Shumovsky-type (KCBS)
systems ($n=5$). But the formula also allows us to deal with the
same experimental paradigm when they create inconsistently connected
systems, due to signaling or contextual biases in experimental design
(see Refs. \cite{DK2014Arxiv1,DK2014CbDArxiv2,DKL2015FooP,KDL2015PRL}
for details).

The contextuality-by-default theory does have precursors in the literature.
The idea that random variables in different contexts are stochastically
unrelated was prominently considered in Refs. \cite{Khr2005,Khr2008,Khr2009,DK2015TNHMP}.
Probabilities of the $\mathsf{eq}$-type with the contextual labeling
of random variables, as defined in (\ref{eq:EQ-type probs}), were
introduced in Refs. \cite{Larsson2002,Simon2001,Winter2014,Svozil}.
The distinguishing feature of the contextuality-by-default theory
is the notion of a maximally connected coupling, which in turn is
based on the idea of comparing maximal couplings for the connections
taken in isolation and those extracted as marginals from the couplings
of the entire system. Contextuality-by-default is a more systematic
and more general theory of contextuality than those proposed previously,
also more readily applicable to experimental data \cite{KDL2015PRL,bacciagaluppi}.

\section{\label{sec:Conceptual-and-Terminological}Conceptual and Terminological
Clarifications}

\subsection{Contextuality and quantum mechanics (QM)}

The notion of (non)contextuality has its origins in logic \cite{Specker1960},
but since the publication of Ref. \cite{Kochen-Specker1967} it has
been widely considered a QM notion. QM indeed provides the only known
to us theoretically justified examples of contextual systems. (Non)contextuality
\emph{per se}, however, is a purely probabilisitic concept, squarely
within the classical, Kolmogorovian probability theory (that includes
the notion of stochastic unrelatedness and that of couplings) \cite{DK2014LNCSQualified,DK2014Scripta,DK2014Advances}.
When contextuality is present in a QM system, QM is relevant to answering
the question of exactly how the noncontextuality conditions in the
system are violated, but it is not relevant to the question of what
these conditions are.

\subsection{Contexts and QM observables}

In particular, the ``objects'' being measured need not be QM observables.
They may very well be questions asked in a poll of public opinion,
and the binary measurements then may be Yes/No answers to these questions.
It is especially important not to confuse being ``measured together''
in the definition of a context with being represented by compatible
(commutative) observables. Thus, in the theory of contextuality in
cyclic systems, $n=4$ is exemplified by the EPR-Bell paradigm, with
Alice's ``objects'' (spins) being $q_{1},q_{3}$ and Bob's $q_{2},q_{4}$.
In each of the contexts $\left(q_{1},q_{2}\right),\ldots,\left(q_{4},q_{1}\right)$,
the two objects are compatible in the trivial sense: any observable
in Alice's Hilbert space $H_{A}$ is compatible with any observable
in Bob's Hilbert space $H_{B}$ because the joint space is the tensor
product $H_{A}\otimes H_{B}$. The case $n=5$ is exemplified by the
KCBS paradigm, where the spins $q_{1},\ldots,q_{5}$ are represented
by observables in three-dimensional Hilbert space. In each of the
five contexts $\left(q_{1},q_{2}\right),\ldots,\left(q_{5},q_{1}\right)$
the observables are compatible in the narrow QM sense: they are commuting
Hermitian operators. The case $n=3$ is exemplified by the Leggett-Garg
paradigm, where three measurements are made at three distinct time
moments, two measurements at a time. The QM representations of the
observables in each of the contexts $\left(q_{1},q_{2}\right),\left(q_{2},q_{3}\right),\left(q_{3},q_{1}\right)$
are generally incompatible (noncommuting) operators. In spite of the
profound differences in the QM structure of these three cyclic systems,
their contextually analysis is precisely the same mathematically,
given by (\ref{eq:measure cyclic}) and (\ref{eq:criterion cyclic}).

\subsection{The meaning of being measured ``together''}

It should be clear from the discussion of the Leggett-Garg paradigm
that ``measuring objects one context at a time'' does not necessarily
have the meaning of chronological simultaneity. Rather one should
think of measurements being grouped and recorded in accordance with
some fixed coupling scheme: if $q$ and $q'$ belong to the same context
$C$, there is an empirical procedure by which observations of $R_{q}^{C}$
are paired with observations of $R_{q}^{C}$. Thus, if the objects
being measured are tests taken by students, and the measurements are
their test scores, the tests are grouped into contexts by the student
who takes them, however they are distributed in time. The grouping
of (potential) observations is in essence what couplings discussed
in Sections \ref{sub:All-possible-couplings} and \ref{sub:Connections-and-their}
do for a set of stochastically unrelated random variables, except
that these couplings do not provide a uniquely (empirically) defined
joint distribution. Rather the probabilistic couplings imposed on
different bunches are part of a purely mathematical procedure that
generally yields an infinity of different joint distributions.

\subsection{The meaning of a noncontextual description}

In the traditional approach to (non)contextuality, where the measurements
are labeled by objects but not by contexts, one can define a noncontextual
description as simply a coupling imposed on the system. For instance,
in the Leggett-Garg paradigm the noncontextual labeling yields three
random variables, $R_{1},R_{2},R_{3}$, with $\left(R_{1},R_{2}\right),\left(R_{2},R_{3}\right),\left(R_{3},R_{1}\right)$
jointly observed. A noncontextual description here is any three-component
random variable $S=\left(S_{1},S_{2},S_{3}\right)$ with $\left(S_{1},S_{2}\right)\sim\left(R_{1},R_{2}\right)$,
$\left(S_{2},S_{3}\right)\sim\left(R_{2},R_{3}\right)$, and $\left(S_{3},S_{1}\right)\sim\left(R_{3},R_{1}\right)$.
The system is contextual if no such description exists.

The situation is different with contextually labeled measurements.
For the Leggett-Garg paradigm we now have six variables grouped into
three stochastically unrelated contexts, $\left(R_{1}^{1},R_{2}^{1}\right),\left(R_{2}^{2},R_{3}^{2}\right),\left(R_{3}^{3},R_{1}^{3}\right)$.
As explained in Section \ref{sub:All-possible-couplings}, such a
system always has a coupling, in this case a sextuple $S$ with $\left(S_{1}^{1},S_{2}^{1}\right)\sim\left(R_{1}^{1},R_{2}^{1}\right)$,
$\left(S_{2}^{2},S_{3}^{2}\right)\sim\left(R_{2}^{2},R_{3}^{2}\right)$,
and $\left(S_{3}^{3},S_{1}^{3}\right)\sim\left(R_{3}^{3},R_{1}^{3}\right)$.
One can call any of these couplings a noncontextual description of
the system. To characterize (non)contextuality then one can use the
term ``maximally noncontextual description'' for any maximally connected
coupling \cite{KDL2015PRL}. Alternatively, one can confine the term
``noncontextual description'' of a system only to maximally connected
couplings for it. With this terminology the definition of a contextual
system in our theory is the same as in the traditional approach: a
system is contextual if it does not have a noncontextual description.
The choice between the two terminological variants will be ultimately
determined by whether couplings other than maximally connected ones
will be found a useful role to play.

\section{\label{sec:Generalizations-Beyond-Binary}Instead of a Conclusion:
Generalizations}

\subsection{Beyond objects and subsets}

Defining a context as a subset of objects measured together \cite{DKL2015FooP,KDL2015PRL}
is less general than defining it by conditions under which certain
objects are measured \cite{DK2014Arxiv1,DK2014CbDArxiv2,deBarros}.
For instance, by the first of these definitions $\left\{ q_{1},q_{2}\right\} $
for a given pair of objects is a single context, while the second
definition allows one to speak of the same pair of objects $q_{1},q_{2}$
forming several different contexts. Thus, if $q_{1},q_{2}$ are two
tests, they can be given in one order or the other, $\left(q_{1},q_{2}\right)$
or $\left(q_{1},q_{2}\right)$. In fact, in all our previous discussion
of cyclic systems we used the notation for ordered pairs, $\left(q_{i},q_{i\oplus1}\right)$
rather than $\left\{ q_{i},q_{i\oplus1}\right\} $. This is inconsequential
for cyclic systems of rank $n\geq3$. For $n=2$, however, the difference
between $\left(q_{1},q_{2}\right)$ and $\left(q_{2},q_{1}\right)$
is critical if $n=2$ is to be a nontrivial system (with the distributions
of the two bunches not identical). It can be shown that the system
can be nontrivial, and $n=2$ is a legitimate value for (\ref{eq:measure cyclic})
and (\ref{eq:criterion cyclic}).

Being formal and mathematically rigorous here makes things simpler.
A context is merely a label (say, superscript) at a random variable
with the convention that identically superscripted variables are ``bunched
together,'' i.e., they are jointly distributed. An object is merely
another label (in our notation, a subscript) that makes all the elements
of a bunch different and indicates which elements from different bunches
should be put together to form a a connection. So if there are six
random variables grouped into three distinct bunches $\left(R_{1}^{1},R_{2}^{1}\right)$,
$\left(R_{1}^{2},R_{2}^{2}\right)$, and $\left(R_{1}^{3},R_{2}^{3}\right)$
and into two connections $\left(R_{1}^{1},R_{1}^{2},R_{1}^{3}\right)$
and $\left(R_{2}^{1},R_{2}^{2},R_{2}^{3}\right)$, we can (but do
not have to) interpret this as three different contexts involving
the same two objects. From mathematical (and perhaps also philosophical)
point of view, measurements grouped into bunches and connections are
more fundamental than objects being measured within contexts.

\subsection{Beyond binary measurements}

How could the definition of a maximally connected coupling be generalized
to arbitrary random variables? A straightforward way to do this is
to extend definition (\ref{eq:EQ-type probs}) for a coupling $T_{q}$
of a connection $\mathfrak{R}_{q}$ as 
\begin{equation}
\mathsf{eq}\left(T_{q}\right)=\Pr\left[T_{q}^{C}=T_{q}^{C'}\textnormal{ for any two }C,C'\in\mathscr{C}\right].
\end{equation}
This is an approach adopted in \cite{DKL2015FooP,KDL2015PRL,deBarros}.
It is based on the following mathematical considerations, derived
from the discussion of maximal couplings in Thorisson's monograph
\cite{Thor} (Section 7 of Chapter 3). 

Given two sigma-additive measures $\mu$ and $\nu$ on the same sigma
algebra, let us write $\mu\leq\nu$ if $\mu\left(E\right)\leq\nu\left(E\right)$
for every measurable set $E$. Let $\mu_{q}^{C}$ be the probability
measure associated with $R_{q}^{C}$. Let $X_{q}$ and $\Sigma_{q}$
be the set of values and sigma algebra associated with $R_{q}^{C}$
(they are assumed the same for all $C$, because otherwise one should
not consider $R_{q}^{C}$ measurements of one and the same object).
For every object $q$, define $\mu_{q}$ as the largest sigma-additive
measure such that $\mu_{q}\leq\mu_{q}^{C}$ for all contexts $C$.
The measure $\mu_{q}$ is the largest in the sense that $\mu'_{q}\leq\mu_{q}$
for any other measure $\mu'_{q}$ such that $\mu_{q}\leq\mu_{q}^{C}$
for all contexts $C$. A theorem proved in Ref. \cite{Thor} (Theorem
7.1) guarantees the existence and uniqueness of $\mu_{q}$, for any
set of probability measures $\left\{ \mu_{q}^{C}\right\} _{C\in\mathscr{C}}$,
whatever the indexing set $\mathscr{C}$. That is, $\mu_{q}$ is uniquely
defined for any connection $\mathfrak{R}_{q}$. Note that $\mu_{q}$
is not generally a probability measure, so $\mu_{q}\left(X_{q}\right)$
can be any number in $\left[0,1\right]$. Let us denote
\begin{equation}
\max\mathsf{eq}\left(\mathfrak{R}_{q}\right)=\mu_{q}\left(X_{q}\right).
\end{equation}
For $\pm1$-measurements $R_{q}^{C}$ this definition specializes
to (\ref{eq:EQ-type probs})-(\ref{eq:meqR_q}).

Consider now a coupling $T_{q}$ for $\mathfrak{R}_{q}$. It is defined
on the product sigma-algebra $\bigotimes_{\mathscr{C}}\Sigma_{q}$
on the product set $\prod_{\mathscr{C}}X_{q}$. An event $E_{q}\in\bigotimes_{\mathscr{C}}\Sigma_{q}$
is called a \emph{coupling event} if $S_{q}\in E_{q}$ implies $T_{q}^{C}=T_{q}^{C'}$
for any two $C,C'\in\mathscr{C}$ (assuming, as always, that both
$C$ and $C'$ involve $q$). It follows from Theorem 7.2 in Ref.
\cite{Thor} that 
\begin{equation}
\Pr\left[T_{q}\in E_{q}\right]\leq\max\mathsf{eq}\left(\mathfrak{R}_{q}\right),
\end{equation}
for any $q$ and any choice of $E_{q}$. Now, it is natural to define
a \emph{maximal coupling} for $\mathfrak{R}_{q}$ as a coupling $T_{q}$
for which $E_{q}$ can be chosen so that
\begin{equation}
\Pr\left[T_{q}\in E_{q}\right]=\max\mathsf{eq}\left(\mathfrak{R}_{q}\right).
\end{equation}
Theorem 7.3 in Ref. \cite{Thor} says that such a maximal coupling
always exists. Note that $E_{q}$ in a maximal coupling can always
be thought of as the largest measurable subset of the diagonal of
the set $\prod_{\mathscr{C}}X_{q}$. 

Having established this generalized notion of a maximal coupling,
the theory of contextuality can now be generalized in a straightforward
fashion. Consider a coupling $S$ for the entire system $\mathfrak{R}$.
The definition of a \emph{maximally connected coupling} remains unchanged:
every marginal $S_{q}$ of a maximally connected coupling $S$ is
a maximal coupling for the corresponding connection $\mathfrak{R}_{q}$.
Our Main Definition could remain unchanged too: a system $\mathfrak{R}$
is \emph{contextual} if and only if no coupling $S$ of this system
is maximally connected. This can be equivalently presented as follows.
For any set $P$ of probability values, let $f(P)$ be a bounded smooth
nonnegative function strictly increasing in all components of $P$.
Thus, for finite systems of random variables $f$ can be chosen as
a sum or average, as in (\ref{eq:ineq. for contextual}). Define
\begin{equation}
\max\mathsf{eq}\left(\mathfrak{R}\right)=\max_{\substack{\textnormal{all couplings}\\
S\textnormal{ for }\mathfrak{R}
}
}f\left(\Pr\left[S_{q}\in E_{q}\right]:q\in Q\right),
\end{equation}
and, if this value exists,
\begin{equation}
\mathsf{cntx}\left(\mathfrak{R}\right)=f\left(\max\mathsf{eq}\left(\mathfrak{R}_{q}\right):q\in Q\right)-\max\mathsf{eq}\left(\mathfrak{R}\right).
\end{equation}
The system is defined as contextual if and only if $\mathsf{cntx}\left(\mathfrak{R}\right)>0$.
We do not know whether $\max\mathsf{eq}\left(\mathfrak{R}\right)$
exists for all possible systems of random variables. If it does not,
however, the definition can be extended by replacing $\max$ with
$\sup$. 

This generalization has to be further explored to determine whether
it is a good generalization, i.e., whether it provides valuable insights,
leads to interesting mathematical developments, and does not yield
non-interpretable results when applied to specific systems of measurements.

\subsubsection*{Acknowledgments.}

This research has been supported by NSF grant SES-1155956 and AFOSR
grant FA9550-14-1-0318.

\end{document}